\documentclass[doublespacing]{elsart}

\usepackage{graphicx}

\usepackage{amssymb}

\begin{document}
\begin{frontmatter}



\title{Design and construction of the Mini-Calorimeter of the AGILE satellite}


\author[iasf-bo]{C. Labanti}
\author[iasf-bo]{M. Marisaldi\corauthref{cor1}}
\ead{marisaldi@iasfbo.inaf.it}
\corauth[cor1]{Corresponding author. Tel.: +39 051 639 8740; fax: +39 051 639 8723}
\author[iasf-bo]{F. Fuschino}
\author[enea-bo]{M. Galli}
\author[iasf-roma]{A. Argan}
\author[iasf-bo]{A. Bulgarelli}
\author[iasf-bo]{G. Di Cocco}
\author[iasf-bo]{F. Gianotti}
\author[iasf-roma]{M. Tavani}
\author[iasf-bo]{M. Trifoglio}

\address[iasf-bo]{INAF-IASF Bologna, Via Gobetti 101, 40129 Bologna, Italy}
\address[enea-bo]{ENEA, Via Martiri di Monte Sole 4, 40129 Bologna, Italy}
\address[iasf-roma]{INAF-IASF Roma, Via del Fosso del Cavaliere 100, 00133 Roma, Italy}

\begin{abstract}
AGILE is a small space mission of the Italian Space Agency (ASI) devoted to gamma-ray and hard-X
astrophysics, successfully launched on April 23 2007. The AGILE Payload is composed of three instruments: a
gamma-ray imager based on a tungsten-silicon tracker (ST), for observations in the gamma ray energy
range 30~MeV - 50~GeV, a Silicon based X-ray detector,
SuperAGILE (SA), for  imaging in the range 18~keV - 60~keV  and a CsI(Tl)
Mini-Calorimeter (MCAL) that detects gamma rays or charged particles energy loss in the range 300~keV - 100~MeV. 
MCAL is composed of 30 CsI(Tl) scintillator bars with photodiode readout at both ends, arranged in
two orthogonal layers. MCAL can work both as a slave of the ST and as an
independent gamma-ray detector for transients and gamma-ray
bursts detection. In this paper a detailed description of MCAL is presented together
with its performance.
\end{abstract}

\end{frontmatter}

\section{Introduction}
AGILE\footnote{The AGILE mission web page: http://agile.rm.iasf.cnr.it/ [2008, April 17] INAF-IASF Roma} \cite{Tavani2008,Tavani2008b} is a small space mission of the Italian Space Agency (ASI) devoted to
astrophysics in the gamma-ray energy range 30~MeV - 50~GeV, with a monitor in the X-ray band 18~keV - 60~keV. The AGILE payload is composed of three instruments: 
a tungsten-silicon tracker (ST) \cite{Barbiellini2001b,Prest2003}, with a large field of view, good time resolution,
sensitivity  and angular resolution; 
a Silicon based X-ray detector,
SuperAGILE (SA) \cite{Feroci2007}, for  imaging in the range 18~keV - 60~keV  and a CsI(Tl)
mini-calorimeter (MCAL) \cite{Celesti2004,Labanti2006,Labanti2006b_proc-Siena-2004} for the detection of gamma-rays or charged particles in the range 300~keV - 200~MeV. 
ST and MCAL form the so called Gamma-Ray Imaging Detector (GRID) for observations in the energy
range 30~MeV - 50~GeV.
The instrument is surrounded by an anticoincidence (AC) system \cite{Perotti2006}, made with
plastic scintillator layers, for the rejection of charged particles and is
completed by the Payload Data Handling Unit (PDHU) \cite{Argan2004}. AGILE was successfully launched on April 23 2007 from Satish Dhawan Space Centre (India) on a PSLV rocket.

The AGILE GRID detection principle is based on the pair production process. The interaction of a high energy photon with a tungsten layer of the silicon tracker originates an electron positron pair whose direction of propagation is sampled by the ST detection panes. ST determines the direction of the incoming radiation,
while MCAL, operating as a slave of ST, measures the energy deposited by particles reaching it. 
MCAL can also work as a stand-alone gamma-ray detector in the
range 300~keV - 100~MeV, with no imaging capabilities, for the detection of
transient events and Gamma Ray Bursts (GRB)  and for evaluation of gamma-ray
background fluctuations. For GRBs in the SuperAGILE field of view, SA will be able to determine the
position of the source and its hard X-ray spectrum, while MCAL will describe the
spectrum above 300~keV and its time variation in correlation with SA. The two instruments are arranged to work together at PDHU level.
Furthermore MCAL produces a broad-band spectrum of the gamma-ray sky (scientific ratemeters) with a refresh rate of 1 sec, for the monitoring of gamma-ray background.  
The architecture of MCAL has been designed to accomplish its different tasks at the same time starting from a single detector system.  

In this paper a detailed description of the MCAL instrument will be provided, from the design solutions adopted to the detector construction and tests. The pre-launch instrument performance will be discussed as well.

\section{Scientific requirements and constraints}
MCAL was conceived as a part of the GRID detector, to complement the ST in the detection of  events, supplying the energy information of the tracked particles.
For this detector its fate is also in its name: mini-calorimeter. AGILE is a small mission with severe constraints concerning payload dimension, weight and power consumption: the overall active payload weight is $120~\mathrm{kg}$, with a power consumption of about $120~\mathrm{W}$. The spacecraft total weight is $350~\mathrm{kg}$.  
Within this frame, just $30~\mathrm{kg}$ and $6~\mathrm{W}$ were allocated for the calorimeter so that the resulting detector could just be mini. 

Apart from the weight and power constraints, few simple concepts had to be followed for the MCAL basic design: its geometrical area should match the ST cross section and it should be capable of determining energy and position of triggered events.
No constraints were put on the complexity of the electronic design of the system and on the number of its functions.
Therefore, since it was not possible to add more scientific instruments to the mission due to payload weight constraints, it  was decided to operate MCAL also to detect gamma-ray transients, the so-called BURST mode, and to monitor the overall low energy gamma-ray background during the orbit.

Concerning the MCAL contribution to GRID operations it was recognised since the early simulations of the AGILE payload that, due to the limited  MCAL thickness (only 1.5 radiation lengths, about $3~\mathrm{cm}$), its efficacy as a calorimeter is restricted to the lower part of the whole GRID energy range. 
For photon energies where MCAL contribution is not conclusive, the energy of the detected photon is evaluated from the scattering angles of the particles moving through the ST detection planes. In this case MCAL can still contribute with a further positional information in the farest location from the gamma-ray interaction.
Moreover, the topological information provided by MCAL (i.e. number of hit detectors, position distribution of events) help the events filtering procedure for background discrimination.

The MCAL design arises from a trade-off between the instrument weight constraint and the efficiency requirements in the detection of high energy gamma-rays.  Its thickness of about 3~cm  gives an acceptable compromise, resulting in a good efficiency for gamma-rays of some MeV. MCAL energy threshold is about 300~keV, resulting as a trade-off itself between the minimum energy threshold requirement and the electronic noise performance as well as low power consumption constraints.

Operating as a GRB monitor, the main goal of MCAL is the detection of fast transients at MeV  energies with microsecond time resolution. In detecting GRBs MCAL is complemented by the SA instrument, which operates at hard X-ray energy and has imaging capabilities in a $68^\circ \times 68^\circ$ field of view.  MCAL can provide only limited information on burst direction, but it has a $4\pi~\mathrm{sr}$ field of view, thus behaving as a true all-sky monitor, even if, for GRBs out of the  SA field of view, a localization from other satellites is required to proceed in spectral analysis (since the detector's response is direction dependent). For GRB detection a dedicated trigger logic must also be included in the system design, in order to trigger on fast rate increases above the background level.

GRBs at MeV  energies have been detected by the BATSE and COMPTEL instruments on-board the Compton Gamma-Ray Observatory (CGRO) during the '90s, and are currently observed by several instruments in space. GRBs at higher energy have been detected by the EGRET instrument on-board CGRO, and GeV emission has been predicted by some models \cite{Meszaros2006} and reported for some bursts \cite{Hurley1994}. There are also many attempts to detect GRB at TeV energy \cite{Atkins2000,Poirier2003},  the most stringent upper limits being currently provided by the MAGIC telescope \cite{Albert2007}. High energy gamma-ray bursts are rare events, which have not been studied in details up to now. For the first time we can study these events with an unprecedented time resolution, better than  $2~\mathrm{\mu s}$, over an energy range spanning six orders of magnitude with all the three main AGILE detectors combined together. This search has recently led to the AGILE detection above 50~MeV of the first GRB after the EGRET era \cite{Giuliani2008}.

\subsection{Monte Carlo simulations}

Extensive simulations were done to verify the scientific performance of the instrument and to optimize its design \cite{Longo2002,Cocco2002b}. A detailed representation of MCAL is included in the Monte Carlo code used for scientific simulations.

   \begin{figure}
   \begin{center}
   \begin{tabular}{c}
   \includegraphics[width=3.5in]{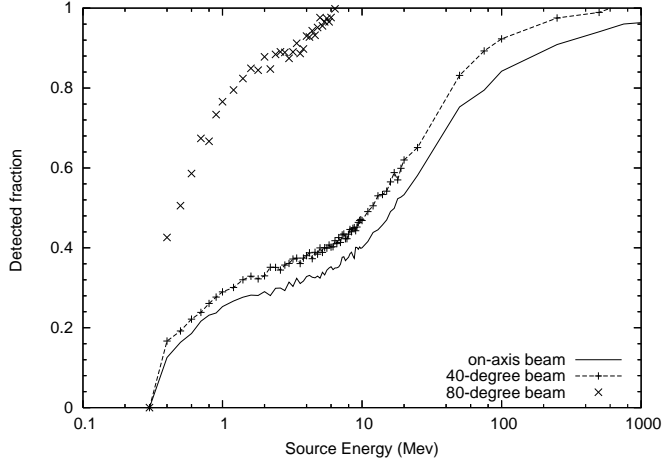}
   \end{tabular}
   \end{center}
   \caption[eff] 
   { \label{fig:eff} Fraction of detected events as a function of energy, for different off-axis angles, from Monte Carlo simulations.
} 
   \end{figure} 

Figure \ref{fig:eff} shows the probability of interaction of photons in MCAL as a function of their energy, from Monte Carlo simulations. It is computed as the ratio between the number of hits observed in MCAL and the number of primary events whose trajectories intersect the MCAL active volume. Since the simulations are carried out considering plane beams hitting the whole AGILE satellite, for high off-axis angles and energies above few MeV the contribution from secondaries originated in the AGILE structure and hitting MCAL becomes important. This is the reason for the efficiency rapidly approaching unity for the 80~degrees off-axis beam shown in figure \ref{fig:eff}.
Below 1~MeV the efficiency trend is very sensitive to the energy threshold that, as it will be explained in section \ref{detection_plane}, is variable on the detector surface. We have assumed here a reasonable but conservative value corresponding to 300 keV at the edge of MCAL bars, where the threshold is lower.

\begin{figure}
   \begin{center}
   \begin{tabular}{c}
   \includegraphics[width=3.5in]{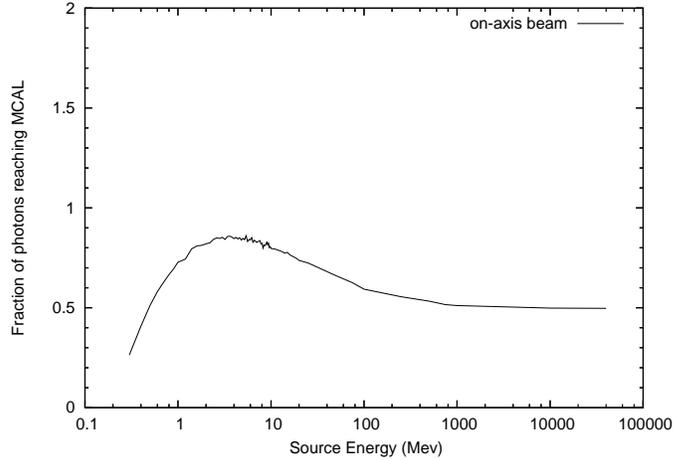}
   \end{tabular}
   \end{center}
   \caption[Frazione] 
   { \label{fig:Frazione} Fraction of photons reaching MCAL without previous interactions as a function of energy. Monte Carlo simulation results for an on-axis parallel plane photon beam.
} 
   \end{figure}

Figure \ref{fig:Frazione} shows the fraction of on-axis photons which reach MCAL prior to any interaction as a function of initial energy. Photons with energy above about $10~\mathrm{MeV}$  mainly interact in the tracker or in the satellite's structural parts, and MCAL detects them  trough secondary products, while photons with energy below 1~MeV can be easily stopped by ST itself.

\begin{figure}
   \begin{center}
   \begin{tabular}{c}
   \includegraphics[width=3.5in]{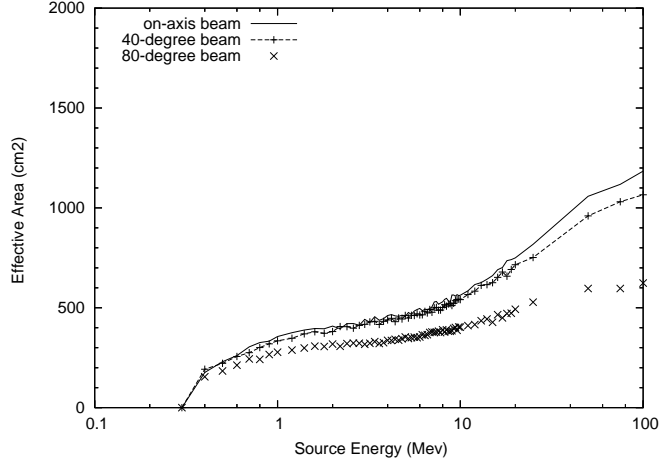}
   \end{tabular}
   \end{center}
   \caption[Aeff] 
   { \label{fig:Aeff} MCAL total effective area as a function of energy.
} 
   \end{figure}

\begin{figure}
   \begin{center}
   \begin{tabular}{c}
   \includegraphics[width=3.5in]{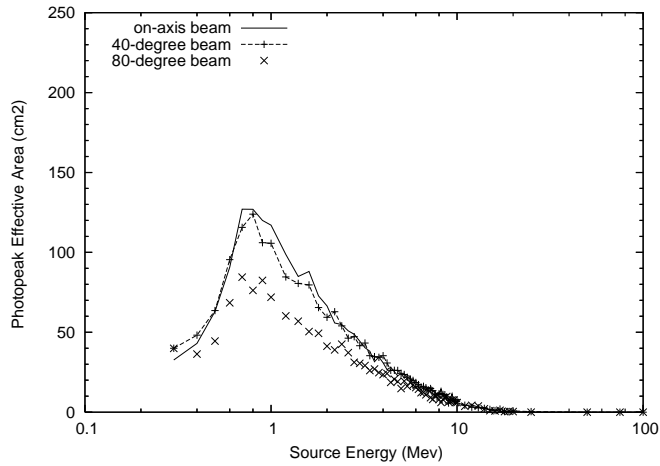}
   \end{tabular}
   \end{center}
   \caption[AeffPK] 
   { \label{fig:AeffPK} MCAL photopeak effective area as a function of energy.
} 
   \end{figure}

Figures \ref{fig:Aeff} and \ref{fig:AeffPK} report the total and photopeak effective area, respectively, as a function of the incident photon energy. Despite the total effective area keeps rising for energies higher than 10~MeV, the photopeak effective area rapidly decreases above a few MeV because of the limited thickness of the detector. 

The effective area plots reported in this section refer to MCAL operated as a self triggering detector, i.e. in the so-called BURST mode, as described in section \ref{opmodes}. MCAL can also work as a slave to the AGILE silicon tracker (GRID mode), but in this case the effective area is determined by the silicon tracker efficiency and trigger criteria, discussed in references \cite{Tavani2008,Tavani2008b}.

\begin{figure}
   \begin{center}
   \begin{tabular}{c}
   \includegraphics[width=3.5in]{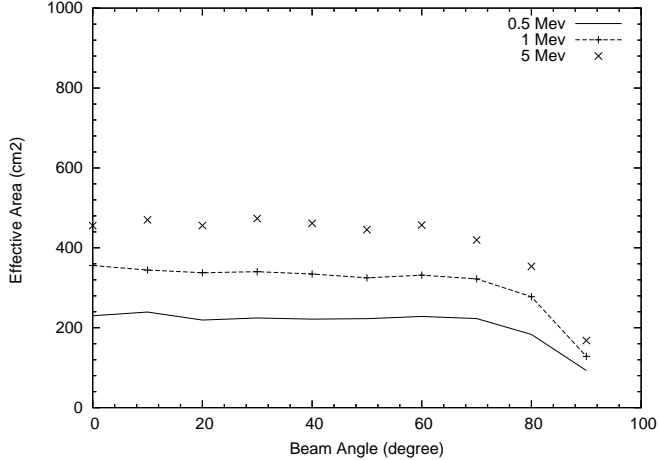}
   \end{tabular}
   \end{center}
   \caption[AeffbyE] 
   { \label{fig:AeffbyE} MCAL total effective area as a function of the angle of incidence.
} 
   \end{figure}

Figure \ref{fig:AeffbyE} shows the total MCAL effective area as a function of the angle of incidence, for different energy values. The effective area remains almost constant for angles between $0^\circ$ (on-axis beam) and $60^\circ$ since the reduction in geometrical area is compensated by an increase in efficiency due to the increase in effective detector thickness. 

Strong efforts have been made to model the expected in-orbit background, as reported in references \cite{Longo2002,Cocco2002b}. Nevertheless, most of the background events in the MCAL energy range are expected to be due to albedo photons and charged particles, which in turn are poorly modeled below 10~MeV. Moreover charged particles can enter MCAL from the spacecraft side without being vetoed by the anticoincidence. Despite these uncertainties, the in-orbit background rate for MCAL in BURST mode is 360~counts/s, in good agreement with our assumptions. The background rate has a 20\% modulation on orbital basis except during the passage through the South Atlantic Anomaly (SAA) where both background rate and dead-time rapidly increase \cite{Marisaldi2008_SPIE}. 

\section{MCAL instrument overview}

MCAL is composed of 30 CsI(Tl) scintillator bars, each one 15x23x375~mm$^3$ in size, arranged in
two orthogonal layers, for a total thickness of 1.5 radiation lengths. In a
bar the readout of the scintillation light is accomplished
by two custom PIN photodiodes (PD) coupled one at each small side of the bar. 
For each bar the PD signals are collected by means of low noise charge
preamplifiers, and then conditioned in the Front End Electronics (FEE). The
circuits have been optimized for best noise performance, fast response,
combined with low power consumption and a wide dynamic range. 
For each bar, the energy and the position of an interacting gamma-ray or ionizing particle can be evaluated combining the signals of the two PDs.

   \begin{figure*}
   \begin{center}
   \begin{tabular}{c}
   \includegraphics[width=6.5in]{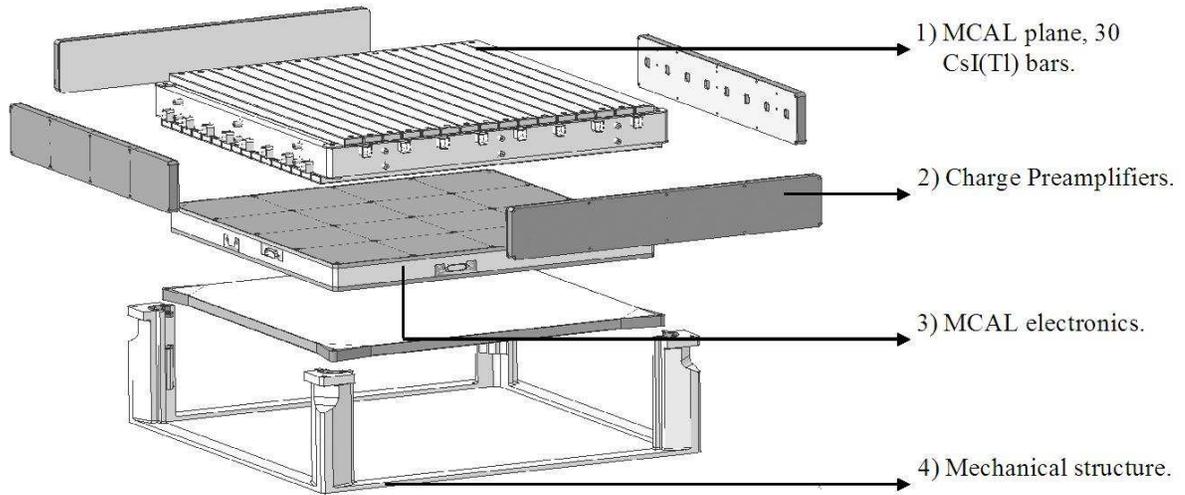}
   \end{tabular}
   \end{center}
   \caption[MCAL_spaccato] 
   { \label{fig:MCAL_spaccato}
Schematic view of the MCAL assembly.} 
   \end{figure*} 

\begin{figure}
   \begin{center}
   \begin{tabular}{c}
   \includegraphics[width=3.5in]{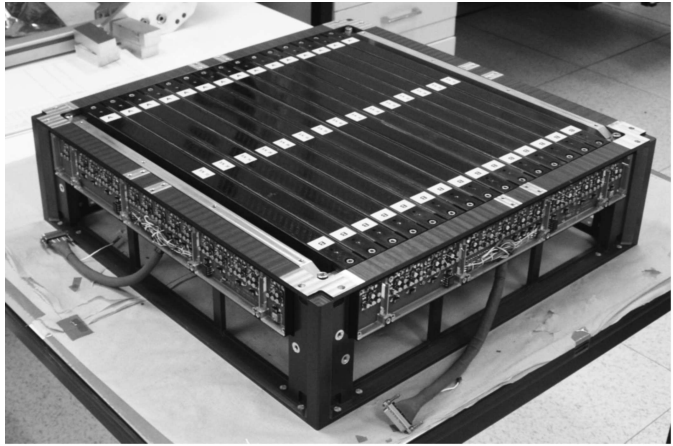}
   \end{tabular}
   \end{center}
   \caption[MCAL_top] 
   { \label{fig:MCAL_top}
Integrated MCAL during test benches, top view. The upper layer of the detection plane and two preamplifiers boards are clearly visible. After payload integration the detection plane faces the ST.} 
   \end{figure}

\begin{figure}
   \begin{center}
   \begin{tabular}{c}
   \includegraphics[width=3.5in]{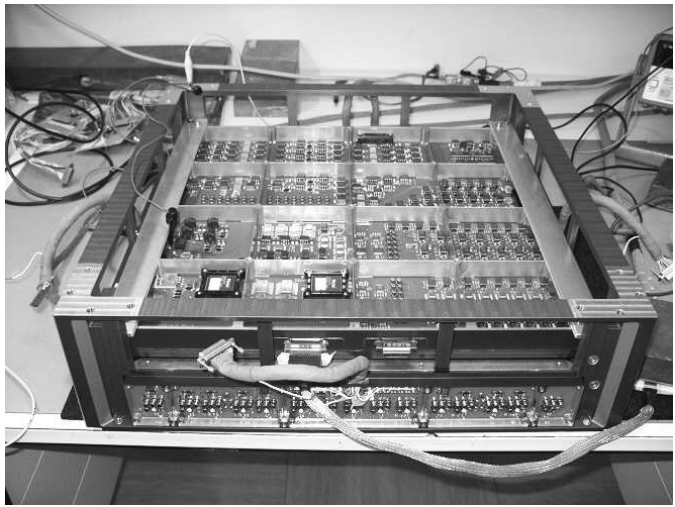}
   \end{tabular}
   \end{center}
   \caption[MCAL_bottom] 
   { \label{fig:MCAL_bottom}
Integrated MCAL during test benches, bottom view. One of the preamplifiers boards and the Front End Electronics are clearly visible. After payload integration the electronics board faces the spacecraft service module.} 
   \end{figure}

Figure \ref{fig:MCAL_spaccato} shows a schematic view of the MCAL assembly. 
The detection plane is hosted in the upper part of MCAL main frame; the
preamplifiers are arranged in four boxes on each side of the detection plane
and at its same level. The preamplifiers are placed as close to the PDs as possible to minimize any stray capacitance. 
Below the detection plane is placed the FEE board that has the same area of the whole detection plane. The overall
mechanical envelope of MCAL constitutes the lower part of the whole AGILE
payload. 
Figures \ref{fig:MCAL_top} and \ref{fig:MCAL_bottom} show the integrated MCAL during test
benches. 

MCAL has been realised by Thales Alenia-Space Italia (TASI), formerly Laben S.p.A., in close collaboration with the AGILE team, who provided the scientific requirement specifications and followed all the design, construction, qualification, test and calibration steps. 
 
\subsection{Operative modes}
\label{opmodes}

MCAL works in two possible operative modes:
\begin{itemize}
\item[-] in GRID mode a trigger issued by the ST starts the collection  of all the detector signals in order to determine the energy and position of particles converted in the tracker and interacting on MCAL;
\item[-] in BURST mode each bar behaves as an independent self-triggering detector and
generates a continuous stream of events in the energy
range 300~keV - 100~MeV. In the data handling system these data are used to
detect impulsive variations in count rates. 
\end{itemize}

Both operative modes can be active at the same time. MCAL setting and operations are managed by tele-commands, while house-keeping data are transmitted to the PDHU to allow the monitoring of the system's health status.
   
\subsection{Scientific ratemeters}
\label{sci_rm}

Due to telemetry limitations BURST data are not sent on ground on a
photon-by-photon basis unless a trigger is issued by a dedicated logic, described in details in \cite{Fuschino2008}. However BURST data are used to build a
broad band energy spectrum (Scientific Ratemeters, SRM) recorded and stored in
telemetry every second. Scientific ratemeters are expected to provide
information on the high energy gamma-ray background in space and
its modulation through orbital phases. Scientific ratemeters are described in details in subsection~\ref{sect:scientific_ratemeters}.

\section{The detection plane}
\label{detection_plane}
The active core of MCAL consists of the CsI(Tl) scintillating bars with two large area PIN PD readout. The PDs have the same footprint of the bar smallest side ($15 \times 23~\mathrm{mm^2}$) with an active area of $256~\mathrm{mm^2}$, a capacitance of about $130~\mathrm{pF}$ and a leakage current of about $1.5~\mathrm{nA}$ at $20~\mathrm{^\circ C}$.
A permanent optical coupling between PDs and CsI is made by means of a clear siliconic resin. This medium has been chosen to realize an elastic bond between two components that exhibit quite a different coefficient for thermal expansion. 
To maximize the light output  and to keep the light attenuation coefficient within an optimal range of values, the bars surfaces are polished and the bars are first wrapped with a reflective coating \cite{Weber2000}  and then with a thin adhesive layer. 
The bars operative temperature range spans from -20 to $+40~\mathrm{^\circ C}$. In such a wide range the thermo-elastic properties of CsI lead to an elongation of about 1~mm on the long bar side. The bar is then arranged inside a carbon fiber structure, about 1~mm thick, that provides rigidity and modularity to the detectors. Cushions realised with soft siliconic resin hold the crystal in place inside the carbon fiber structure allowing thermal expansion.

A detector bar is characterized by:
\begin{itemize}
\item[-] the charge per unit energy produced at PD level by a 
  gamma-ray interacting in the crystal at a defined distance from the PD. 
This quantity will be referred to as signal output, and expressed in $e^-/\mathrm{keV}$.
\item[-] the relation governing the signal output as a function of the
  distance of interaction from the PD. This relation will be referred to as light attenuation law.
\end{itemize}
The MCAL bar surfaces are polished and wrapped in such a way to exhibit an
exponential light attenuation law to a good approximation level. The PDs signal $U(x)$, expressed in $e^-$, as a function of the
distance of interaction $x$ from the PD can then be described by the relation: 

	\begin{equation}
	\label{eq:bar_signal}
U(x) = E U_0 e^{-\alpha x}
	\end{equation}
where $E$ is the energy released in the bar, $U_0$ is the extrapolated signal
output for interactions at the PD edge and $\alpha$ is defined as the light
attenuation coefficient.
The evaluation of energy and position of interaction in a bar comes straightforward from this relation, as shown in appendix \ref{reconstruction}.
The accuracy of these evaluations depends on the signal amplitude, and on both statistical and electronic noise. The same parameters also affect the minimum detectable energy.
Actually the energy threshold is not a constant for a given bar but depends also on the position of interaction along the bar. The edges of the bars, near the PDs, are more sensitive and exhibit a lower energy threshold, while the central part of the bar is less sensitive. This is due to the exponential behaviour of the light output and the discriminator logic that acts on the sum of the signals of the two PDs. According to the measured bar parameters, the difference in energy threshold between the edges and the central part of a bar is of order 30-40~keV. This behaviour has been properly accounted for in Monte Carlo simulations.

\subsection{Bars stand alone characterization}

The parameters $U_0$ and $\alpha$ have been measured
independently for each bar, before integration into the MCAL flight model,
exposing each bar to a collimated $^{22}$Na source at different positions.
It has been found that an exponential model for the light
attenuation law is quite a good representation for the bars along all the
bar length but the first few centimeters near the PDs, where border effects are responsible for a deviation from the above model up to 5-7\%. 

\begin{figure}
   \begin{center}
   \includegraphics[width=3.2in]{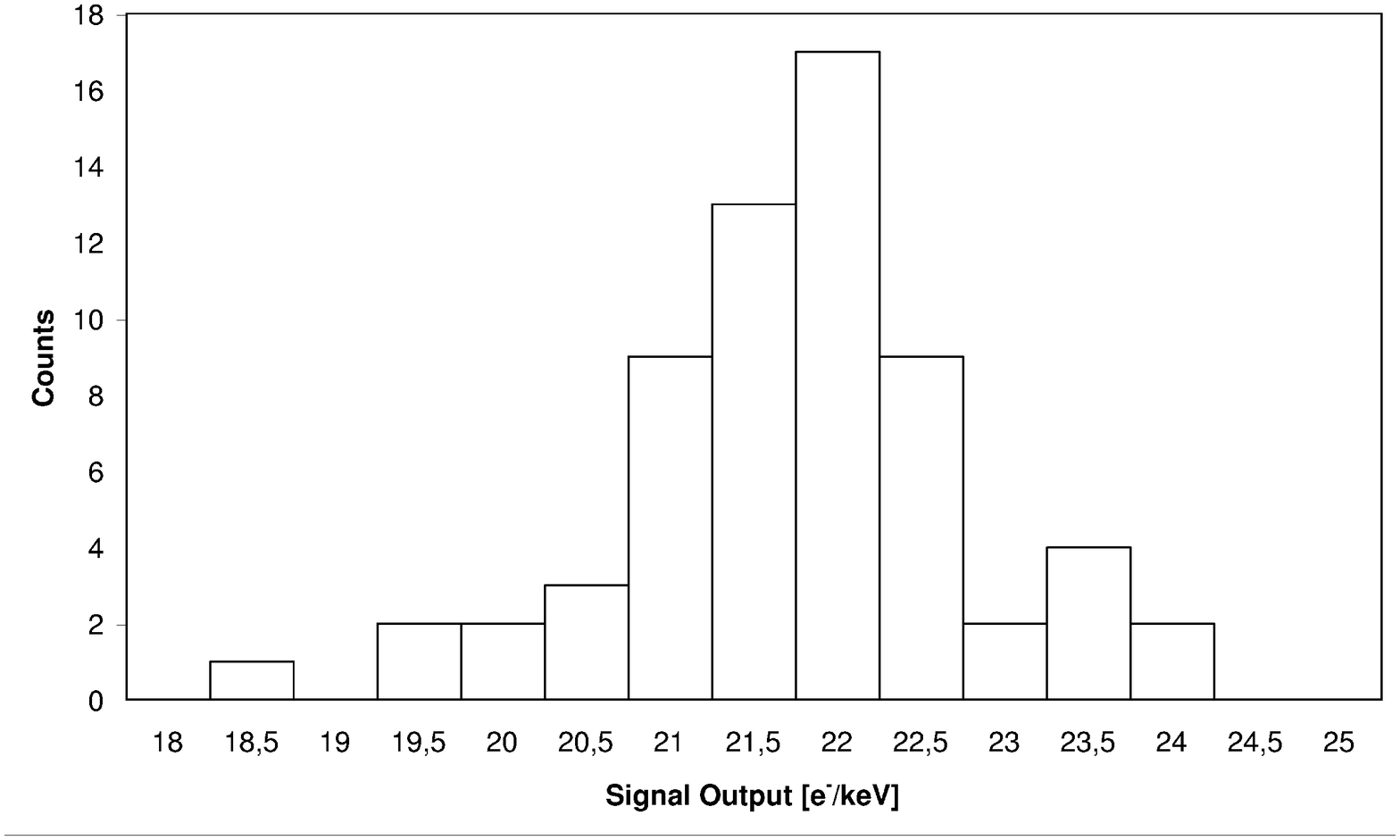}
   \includegraphics[width=3.2in]{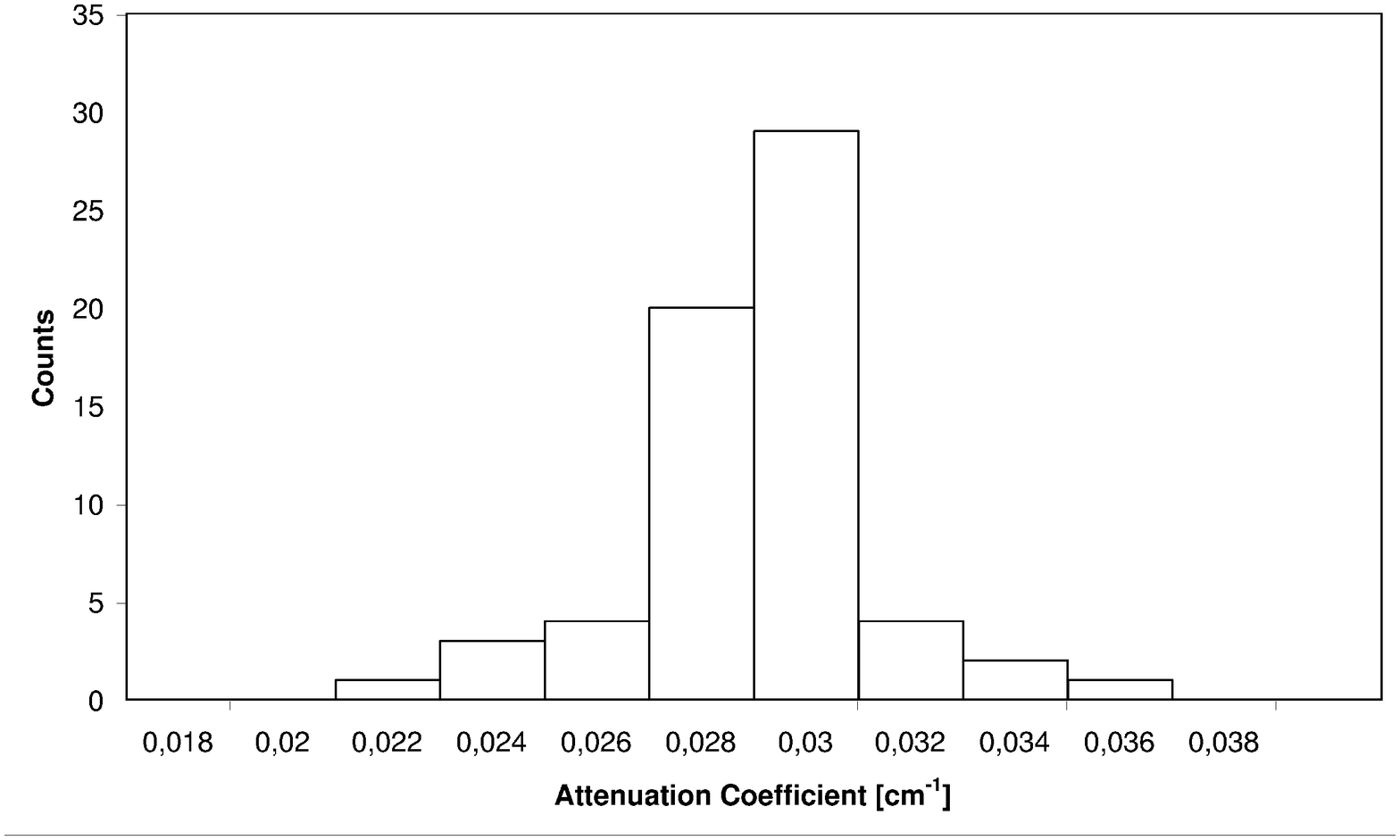}
   \end{center}
   \caption[E_pos_reconstruction] 
   { \label{fig:bar_parameters} 
Bar parameters distribution for the flight detectors. Top panel: signal
   output extrapolated at PD edge $U_0$. Bottom panel: attenuation coefficient
   $\alpha$.}
   \end{figure}

Among the whole set of 32 bars (30 flight detectors plus 2 spares), the average
value for $U_0$ is $21~e^-/\mathrm{keV}$ with a standard deviation of
$1~e^-/\mathrm{keV}$, while the average value for $\alpha$ is
0.028~cm$^{-1}$ with a standard deviation of 0.002~cm$^{-1}$. 
Figure \ref{fig:bar_parameters} shows the distribution of $U_0$ and $\alpha$ for the flight detectors.

The best detectors, i.e. those with higher signal output
and lower attenuation coefficient, have been placed at the external edges of the MCAL planes. This
choice is based on Monte Carlo simulations and is aimed at maximizing the MCAL efficiency in both GRID and BURST modes of operation.

\subsection{Front-end electronics}
MCAL Front End Electronics is physically divided in two different parts:
the PD charge preamplifiers are mounted in four boxes very close to the PDs,
while all the rest of the electronics is placed on a single board, 415x415~mm$^2$ in size, which is placed below the detectors and hosts about 5000 components.
MCAL FEE elaborates the signals of the 60 PDs producing the data stream for both GRID and BURST modes, with just 4.5~W power consumption (including about 1~W consumed by the preamplifiers). 
The analogue chain of each PD is common for GRID and BURST operations up to
the shaping amplifier stage, with a shaping time of 3~$\mu$s to cope with CsI(Tl) scintillation light decay time. 
The electronic noise of each analogue chain is about 1000~$e^-$~rms.
The amplified signals are then sent to the GRID and BURST branches for further processing. 

For GRID operations the signals are sent to a stretch and hold circuit whose
operation is commanded by the ST with a proper delay of a few $\mu$s that
accounts for the different timing response of the detectors. In GRID mode all
the PD signals are stretched and A/D converted via a multiplexer
and a single ADC system.  A de-randomizing FIFO buffer is placed between the GRID
branch and the PDHU. The dead time of the whole operation is about 65~$\mu$s.
In GRID mode, only in one case MCAL can trigger the conversion of
signals. This happens when an energy deposit greater than a threshold,
selectable by tele-commands between 50 and 100~MeV, is detected on the whole MCAL. Such a high energy deposit is expected
to be caused by particle back-splash following a high energy gamma-ray
interaction. Such particles could trigger also the veto system and, without the MCAL trigger, the event would be lost. This MCAL trigger is achieved shaping the sum of all the PD signals with $1~\mathrm{\mu s}$ shaping time and with a fast threshold
discriminator. This operation can be enabled via telecommand.

For BURST operations the amplified signals from the two PDs of one bar are summed and sent to a programmable threshold discriminator circuit. The signal discrimination is made in two steps: a simple threshold level circuit is used to enable a zero-crossing discriminator. This latter stage has been added to increase the timing accuracy, since burst events are time tagged on the logical signal issued by the discriminator. 
In the BURST branch the detectors act independently from each other. The
triggers generated by signals above threshold are sent to an FPGA, that
verifies the logical conditions for processing the signal, for example,
looking at the AC status. Only signals from the two PDs of a bar
generating a valid trigger will be stretched and then passed to a 12-bit ADC by means of a multiplexer. 
An event is completely described adding to the
two PD signal levels the address of the bar and a time mark. The time
resolution for BURST events is determined by the discriminator jitter
characteristics and is lower than 2~$\mu$s. In case of coincident events in more than one bar a sparse readout of the triggered bars is applied to minimize the dead time. 
BURST data are stored in a de-randomizing FIFO buffer and then sent
to the PDHU where they are continuously processed for the detection of fast
transients and the generation of scientific ratemeters.

MCAL FEE includes also ancillary functions for its setting and monitoring, as
house-keeping data generation. These data include voltages, temperatures and several ratemeters to keep track of the trigger rate of the various discriminators.

Table \ref{tab0} shows the main characteristics of the detector.

\begin{table}
\caption{\label{tab0} MCAL physical and engineering characteristics}
\centering
\begin{tabular}{l l}
\hline 
Detector property        &  Measured value \\
\hline 
Active detector weight           &   20~kg \\
On-axis geometrical area         &   $1400~\mathrm{cm^2}$ \\
Number of independent detectors  &   30 \\
Single detector dimensions       &   $15 \times 23 \times 375~\mathrm{mm^3}$ \\
Detector light output            &   $21 \pm 1~e^-/\mathrm{keV}$ \\
Detector attenuation coefficient &   $0.028 \pm 0.002~\mathrm{cm^{-1}}$ \\
Total power consumption          &   $<6~\mathrm{W}$ \\
Electronic noise / channel       &   $<1000~e^-~\mathrm{RMS}$ \\
\hline
\end{tabular}
\end{table}

\subsection{Anticoincidence setting}

The goal of the anticoincidence system is the in-flight rejection of charged particle background.
Its efficiency has been evaluated to be greater than 0.9999 by means of tests with radioactive sources on the plastic scintillator panels before payload integration, and with muon tracks detected in the silicon tracker on ground at integrated payload level. The fast-rising AC veto signal must then be stretched and delayed to cope with the different timing of the PD signals. The delay and width of the AC shaped pulse can be set by tele-commands and have been adjusted in order maximize the rejection efficiency on MCAL.

\section{Interface with the Payload Data-Handling Unit}

\subsection{Data types}
\label{datatypes}
In the PDHU MCAL data are formatted into different Telemetry Packet types, according to the operative mode they refer to, as described in the following list:\\

\begin{itemize}
\item[-] \emph{GRID Observation} data: all MCAL GRID chains are read out when a trigger is issued by the ST. Only bars showing an energy deposit higher than a proper threshold (about 1~MeV) are selected. For each bar above threshold the two A/D converted PD signals are saved in the telemetry packet together with the tracker information. The event timestamp is provided by the ST. This data type contains the most relevant data for gamma-ray analysis with the ST. A strict trigger policy is applied for on-board background rejection.

\item[-] \emph{GRID Physical Calibration} data: all MCAL GRID chains are read out when a trigger is issued by the ST. The A/D converted PDs signals from all the 30 MCAL bars are saved in the telemetry packet together with the tracker information. The event timestamp is provided by the ST. For this data type the trigger selection policy is less strict than in the previous case so that charged particle tracks can be recorded. This data type is for calibration purposes.

\item[-] \emph{MCAL BURST Physical Calibration} data: every event self-triggered by MCAL in the BURST branch is saved in this packet type. For each event the A/D converted PD signals from all the triggered bars are saved. The event timestamp is provided by the PDHU when it receives a logical signal from the MCAL BURST branch discriminator. This data type is  for calibration purposes.

\item[-] \emph{MCAL BURST Observation} data: after a burst trigger is issued by the Burst Search logic, every event self-triggered by MCAL in the BURST branch is saved in this packet type. The event structure is the same as that described above, but only events included in the time interval defined by the Burst Search logic are saved. Additional information concerning the Burst Search status at trigger time are recorded. This data type is used during in-orbit nominal operations.

\item[-] \emph{MCAL BURST Electrical Calibration} data: the MCAL preamplifiers are stimulated with an integrated pulse generator and every event self-triggered by MCAL in the BURST branch is saved in this packet type. The event structure is the same as that described above. Typically four pulses of different amplitude are used to span all the MCAL dynamic range, and at least 500 events for each amplitude are recorded. This data type is used for BURST mode calibration purposes.

\item[-] \emph{Zombie Report} data: every event self-triggered by MCAL in the BURST branch during a GRID event readout is saved in this packet type. The event structure is the same as that described above. The name of this data type derives from the consideration that stored events are produced in the independent MCAL BURST branch, that is still 'alive' during the relatively long 'dead' time of the GRID readout. 

\item[-] \emph{Scientific Ratemeters} data: a collection of ratemeters from different AGILE instruments. As concerns MCAL, the 11~bands spectra for both detection layers are recorded with a 1.024~s time resolution, as described in section \ref{sci_rm}.

\item[-] \emph{House-keeping} data: a collection of analog and digital house-keeping signals from different AGILE instruments, recorded every 16~s and used to monitor the health of the system. As concerns MCAL, the detectors and electronics temperature, as well as the sensible currents, voltages and trigger ratemeters are stored.
\end{itemize}

\section{Environmental and qualification tests}
MCAL was built in accordance with the strict requirements of a space-born instrument.
Since the early stages of detector design, representative samples of the main and most critical elements of the system were tested with environmental cycles that stressed their mechanical and electrical properties. 

\begin{figure}
   \begin{center}
   \begin{tabular}{c}
   \includegraphics[width=3.5in]{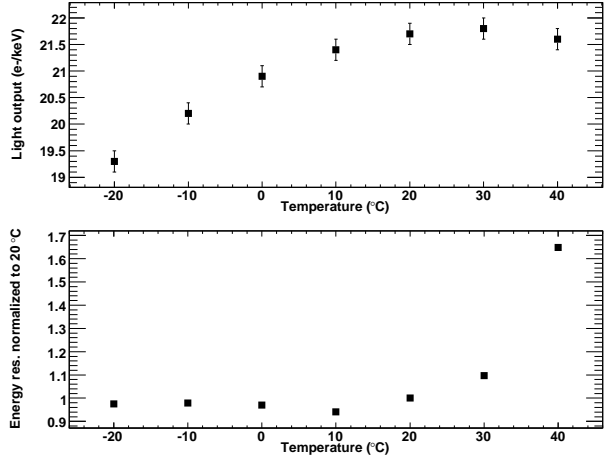}
   \end{tabular}
   \end{center}
   \caption[bar_LY] 
   { \label{bar_LY}
     Qualification test results on a single bar detector. Signal output (top panel) and energy resolution FWHM at 1275~keV (bottom panel), normalized to the value at 20~$^\circ$C, as a function of operating temperature.
   } 
   \end{figure}

These tests were also useful to characterise, for example, the bar detector response vs temperature.
MCAL operative temperature range is  -20~+30~$^\circ$C, while the allowed non operative range is  -20~+40~$^\circ$C. 
In the non operative range the length of a bar made of a 'soft' material like CsI varies about 1~mm, the scintillation light production and the noise of the electronics can change notably too.
Before the production of all the flight detectors, a prototype was tested in the temperature range
 -20~+40~$^\circ$C, with 10~$^\circ$C steps. At each temperature step, spectra of a collimated $\mathrm{^{22}Na}$ source
have been collected at various positions along the bar to derive the parameters described in section \ref{detection_plane}. 
The same tests were also performed after stressing the prototype with several temperature cycles. The results have shown that the detector's performance are stable against thermal stresses in the operative and non-operative range of the instrument demonstrating the reliability of the chosen technical approach.
Figure \ref{bar_LY} shows the signal output and the energy resolution FWHM at 1275~keV as a function of temperature, normalized to the value at 20~$^\circ$C, for one of the test detectors. Below 20~$^\circ$C the energy resolution is almost constant, indicating that the dominant term in the electronic noise in this temperature range is due to the PD capacitance. The degradation of energy resolution at temperatures above 30~$^\circ$C means that the leakage current term contribution to the electronic noise becomes dominant.

MCAL as a whole system was tested on harsh environmental conditions once it was assembled in the payload and in the satellite. At satellite level it  was characterised before and after thermo-vacuum, electro-magnetic compatibility, vibration and acoustic tests. A set of performance and functional tests have been executed after each environmental test, and after transportation to the launch site, in order to identify any possible change with respect to the pre-test situation. Typical monitored parameters were offset, gain and electronic noise of the GRID branch, for all 30 detectors, by means of measurements of cosmic muons. No significant modification of these parameters due to the environmental tests have been observed.

\section{On-ground calibration}

Once it was fully assembled MCAL underwent several calibration sessions. 

\begin{itemize}
\item[-] \emph{MCAL stand-alone calibration:} calibration performed at instrument level, prior to integration into the AGILE payload. Data acquisition was performed by means of a dedicated test equipment simulating the PDHU. 
Using data produced in BURST mode the parameters described in section \ref{detection_plane} and the threshold level were measured for every bar using standard radioactive sources. 
To verify the proper functionality of the GRID chain an external signal simulating the trigger from the silicon tracker was needed, in coincidence with the MCAL event. This has been obtained both using natural muons triggered by a plastic scintillator detector placed above MCAL, and using tagged 511~keV gamma-rays from a $\mathrm{^{22}Na}$ source. Using the same triggering system it was also possible to properly set the anticoincidence timing parameters and evaluate its efficiency.

\item[-] \emph{Payload level calibration:} calibration performed at the DA$\mathrm{\Phi}$NE accelerator Beam Test Facility (BTF) \cite{Mazzitelli2003_NIM515} of the INFN Laboratori Nazionali di Frascati, using a custom photon tagging system. This  session was mainly dedicated to the GRID calibration. 

\item[-] \emph{Satellite level calibration:} performed after payload integration into the spacecraft structure. This session was performed exposing MCAL to an un-collimated $\mathrm{^{22}Na}$ radioactive source placed at different positions with respect to the satellite axis in order to evaluate the MCAL efficiency and the overall contribution of the spacecraft volumes to the detector response. At this level, also the MCAL capability to detect gamma-ray transients was tested. For this purpose a dedicated burst-simulator device was realised, as described in \cite{Fuschino2008}. The basic principle of this device consists in moving a radioactive source in front of a collimator illuminating MCAL to create an artificial rate increase; the speed of the source determining the duration and rise time of the Burst.
\end{itemize}

\section{On-ground performance}

Table \ref{tab1} shows the main MCAL scientific performance.

\begin{table}
\caption{\label{tab1} MCAL main scientific performance}
\centering
\begin{tabular}{l l}
\hline 
Instrument property  &  Measured value \\
\hline 
Energy resolution        &   14\% FWHM at 1.275 MeV \\
Position resolution      &   18~mm at 1.275 MeV \\
                         &    7~mm at 11 MeV \\
Timing accuracy	         &    2~$\mu$s \\
BURST mode energy range  &    0.33 - 110 MeV (near PDs) for each bar   \\
GRID mode energy range   &    1 - 100 MeV (near PDs) for each bar \\
Effective area           &    $\sim$200 cm$^2$ at 0.4 MeV on axis \\
                         &    $\sim$300 cm$^2$ at 1 MeV on axis \\
Field of view            &    $\sim 4 \pi$ sr, non imaging \\
\hline
\end{tabular}
\end{table}

\subsection{Electronic noise evaluation}

Electronic noise estimation for all the acquisition chains was obtained using light output and gain data, together with pedestal peaks width obtained from GRID Physical Calibration data. Since in this data type for every ST trigger all bars data are stored, as described in section \ref{datatypes}, every PD spectrum shows a pedestal peak corresponding to electronic noise sampling. Fitting this pedestal peak with a Gaussian leads to an estimation of the electronic noise of the corresponding electronic chain, according to the following equation 

	\begin{equation}
	\label{eq:noise}
{\sigma}_{el} = \frac{U_0}{u_0} \sqrt{{\sigma}^{2}_{tot} - 1}
	\end{equation}
where $U_0$ is the signal output at PD edge as described in section \ref{detection_plane} expressed in $e^-$/keV, $u_0$ is the gain of the electronic chain under test, expressed in ADC~channels/keV, as described in Appendix \ref{reconstruction}, and $\sigma_{tot}$ is the measured pedestal standard deviation expressed in ADC~channels. The $-1$ term inside the square root accounts for the intrinsic ADC conversion error. 
With this method the average electronic noise is 920~electrons rms with a standard deviation of 70~electrons rms.

This procedure has also been used as a check of the health status of the electronics during qualification and environmental tests.

\subsection{Energy resolution}

\begin{figure}
   \begin{center}
   \begin{tabular}{c}
   \includegraphics[width=3.5in]{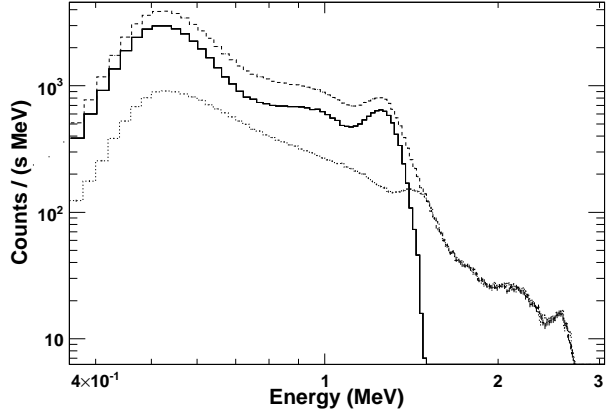}
   \end{tabular}
   \end{center}
   \caption[11680_bkg-sub] 
   { \label{fig:11680_bkg-sub}
MCAL $\mathrm{^{22}Na}$ spectrum obtained in BURST mode during on-ground calibrations. Only single events have been selected. Dashed line: measured spectrum; dotted line: background spectrum; straight line: background-subtracted energy spectrum.} 
   \end{figure}

A typical MCAL bar exhibits an energy resolution of about 14\%~FWHM at 1.275~MeV for un-collimated events.
After the single bar energy calibration, the energy resolution in BURST mode has been evaluated for MCAL considered as a single detector. Figure \ref{fig:11680_bkg-sub} shows the spectra obtained with an un-collimated $\mathrm{^{22}Na}$ radioactive source placed on top of the payload during on-ground calibrations at integrated satellite level. The background spectrum and the corresponding background-subtracted energy spectrum are also shown. Only single events, which nevertheless account for more than 90\% of the events, have been selected. The 1275~keV peak in the background-subtracted spectrum can be fitted with a Gaussian with 88~keV standard deviation leading to an energy resolution $\frac{\Delta E}{E}=18\%$~FWHM. This value is higher than that obtained for single bars, accounting for differences in bars equalization and operative temperature. 
  
\begin{figure}
   \begin{center}
   \begin{tabular}{c}
   \includegraphics[width=3.5in]{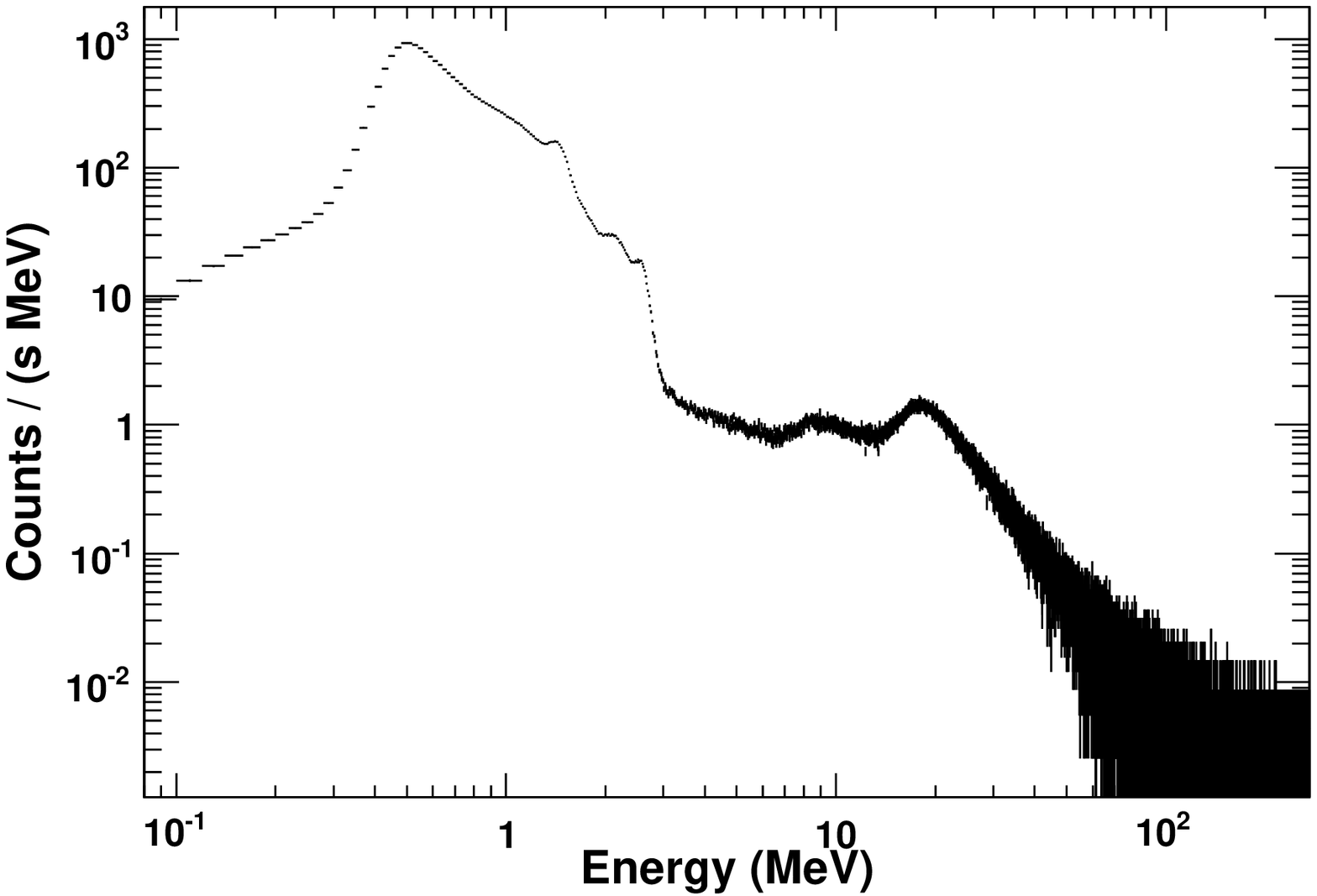}
   \end{tabular}
   \end{center}
   \caption[6152_spectrum] 
   { \label{fig:6152_spectrum}
MCAL background count spectrum obtained in BURST mode during satellite integration. Spectral features due to radioactive isotopes in the environment and cosmic muons can be observed. Measurement duration is 11500~s.} 
   \end{figure}

Figure \ref{fig:6152_spectrum} shows a background spectrum recorded with MCAL during satellite integration. This spectrum was obtained considering MCAL as a single detector, i.e. summing together the contributions from different bars triggering at the same time. The low energy part of the spectrum is dominated by the $^{40}$K 1.460~MeV peak and by other features due to natural radioactivity. At higher energies the spectrum is dominated by two broad peaks at about 10 and 20~MeV. These peaks are due to cosmic muons crossing one or two MCAL planes, respectively. 10~MeV is about the expected energy loss by a minimum ionizing particle (MIP) in 1.5~cm of CsI. 

Since energy resolution depends on the detector and the first amplifying stage but not on the operative mode, the reported values should be the same for both GRID and BURST operative modes.

\subsection{Position resolution}

\begin{figure}
   \begin{center}
   \begin{tabular}{c}
   \includegraphics[width=3.5in]{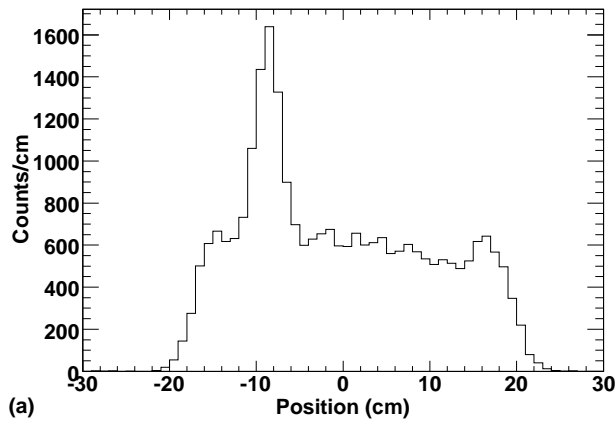}
   \end{tabular}
   \end{center}
   \caption[336_X10_pos_sel-E] 
   { \label{fig:336_X10_pos_sel-E}
Position distribution along an MCAL bar (top plane, bar n.10) for events with energy greater than 1~MeV. Measurement performed in BURST mode with a $\mathrm{^{22}Na}$ collimated source during MCAL stand-alone calibrations.} 
   \end{figure}

For BURST mode the position resolution was evaluated using the collimated $\mathrm{^{22}Na}$ measurements performed for stand-alone calibration. Figure \ref{fig:336_X10_pos_sel-E} shows the position distribution for one of the MCAL bars, for events with energy greater than 1~MeV. The plateau is due to the background radiation, mainly related to $\mathrm{^{40}K}$ decay at these energies, while the peak centered at -8.6~cm with 1.3~cm standard deviation is due to the $\mathrm{^{22}Na}$ line at 1.275~MeV. The collimator opening was 2~mm and the source beam was hitting at position -8.75, according to the reference frame described in Appendix \ref{reconstruction}. At this energy typical standard deviation values span between 1.3 and 1.8~cm depending on the bar and position.

\begin{figure*}  
\begin{center}
\begin{tabular}{c}
\includegraphics[width=6.5in]{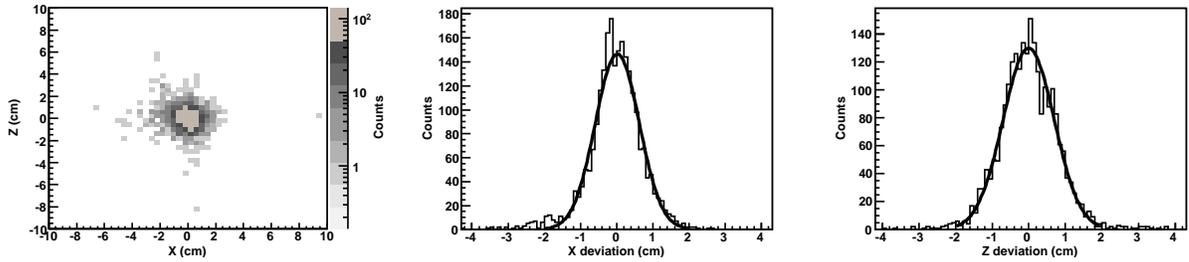}  
\end{tabular}
\end{center}
\caption[grid_pos_reconstruction]
{\label{fig:grid_pos_reconstruction}
Deviation from extrapolated and reconstructed positions for muon tracks in GRID mode. Left panel: 2D deviation distribution. Central panel: deviation distribution along X direction. Right panel: deviation distribution along Z direction. Gaussian fits exhibit a 0.7~cm standard deviation in both directions. Measurements taken on ground during satellite integration.} 
\end{figure*}

In GRID mode position resolution can be estimated using cosmic ray muon tracks. Muon tracks can be easily selected in a GRID dataset requiring straight trajectories with low scattering angles and a track extended over most of the ST detection planes. Muons trajectories can then be extrapolated to MCAL to obtain the expected interaction position. After gain calibration, from the MCAL data the reconstructed position of interaction was calculated and compared to the extrapolated value. A good agreement was obtained, the distribution of the deviation with respect to the expected value having a 0.7~cm standard deviation, as shown in figure~\ref{fig:grid_pos_reconstruction}. After fine tuning of calibration parameters an even better position resolution is expected. This procedure will be used also during flight operations for MCAL gain calibration in GRID mode, using cosmic rays.

Despite the limited positional capabilities of MCAL, a rough estimation of the incoming direction for bright transients observed in BURST mode is expected. That will be accomplished comparing the bars position distribution after background subtraction, taking advantage of the shadow cast by the silicon tracker on MCAL, particularly significant in the few hundreds keV energy range where most of the GRBs have their maximum luminosity. Also the net count rate difference between the two detection layers will help to figure out whether the angle between the burst incoming direction and the AGILE pointing direction is less than $90^\circ$ or not. Since the most valuable information would come from a coincident detection of a transient both in MCAL and in the silicon tracker, even this coarse positional information would provide a preliminary ranking of the event, indicating whether a detailed search in the GRID data is worth doing or not.

\subsection{Timing accuracy}

Timing resolution was estimated using the 50~Hz pulsed beam of the $\mathrm{DA{\Phi}NE - BTF}$ facility during integrated payload calibration activities. Since an absolute time reference was not available, the following procedure has been used. A BURST Physical Calibration dataset acquired during beam operations has been filtered to select only events with energy greater than 1~MeV. The distribution of the time difference between successive events shows an evident periodic behaviour, 20-ms separated peaks, clearly due to the beam pulses. Events on each peak are likely due to the beam. Each of these peaks can be fitted with a Gaussian distribution with a standard deviation of 2.8~$\mu$s. Since this value includes both the possible fluctuations of the 50~Hz signal and the MCAL timing resolution, the latter term must be lower than 2.8~$\mu$s.

\subsection{Scientific ratemeters generation}
\label{sect:scientific_ratemeters}
\begin{figure}
   \begin{center}
   \begin{tabular}{c}
   \includegraphics[width=3.5in]{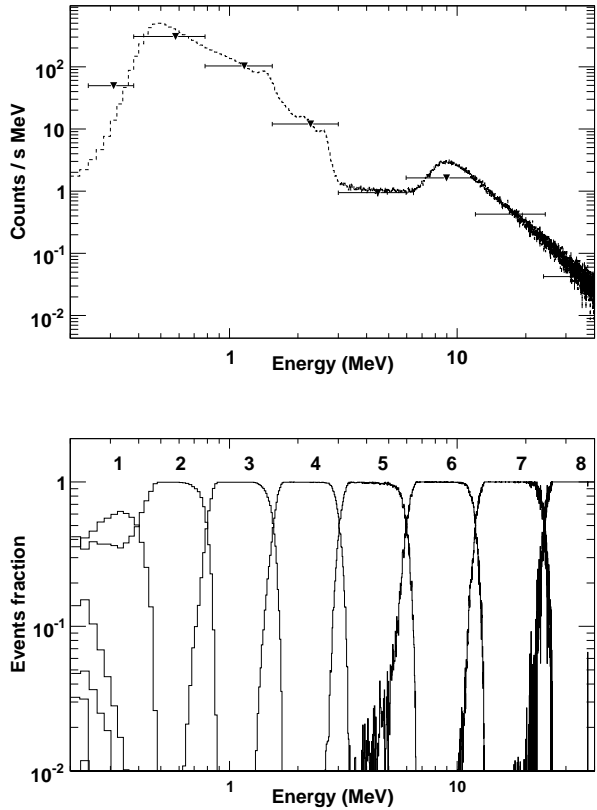}
   \end{tabular}
   \end{center}
   \caption[6152_X_sciRM_bands] 
   { \label{fig:6152_X_sciRM_bands}
Top panel: scientific ratemeters spectrum (triangles) and photon-by-photon reconstructed energy spectrum (dashed line) for the MCAL top detection layer obtained during satellite integration. Bottom panel: fraction of events assigned to a SRM band as a function of the reconstructed energy; numbers indicate the SRM band number.} 
   \end{figure}

For scientific ratemeters (SRM) generation, since a proper energy reconstruction based on
calibration parameters would be an excessive load for the PDHU, an
approximated algorithm was used. For each triggered bar the 
event's energy is evaluated by summing the A/D converted signals of the two PDs
and subtracting the corresponding offsets. If a multiple event has occurred,
the energies evaluated for all the triggered bars are summed together. Although this algorithm for energy estimation is quite rough, it has been
verified that energy is evaluated within 20\% from its true value, thanks to
the good bars transparency and the limited spread in the bars and electronics
parameters. SRM are organized in 11 bands for each of the two MCAL detection layers. Each event is
assigned to one of the energy bands according to the most significant bit of
the energy binary word. Though this assignment policy is straightforward from the
computational point of view, it also defines the energy bands, as each of
them is twice the extension of the previous one. The first useful band includes
events between 0.18 and 0.35~MeV, the second band between 0.35 and 0.7~MeV, the third is between 0.7 and 1.4~MeV and so
on until the last band which includes events with energy higher than
180~MeV. Fig.\ref{fig:6152_X_sciRM_bands} shows a SRM spectrum for the MCAL upper
detection layer recorded during satellite integration phase in January 2006 and its comparison with the count spectrum reconstructed from the corresponding BURST Physical Calibration dataset. The bottom panel in Fig.\ref{fig:6152_X_sciRM_bands} shows the fraction of events assigned to a SRM band as a function of the reconstructed energy. It can be seen that the SRM bands overlap each other. This crosstalk is due to the rough algorithm and to differences in the bars parameters, and is particularly evident at low energy, where small errors in the offset values may significantly alter the energy reconstruction.
Despite these limitations SRM will be a useful tool to monitor the gamma-ray background evolution during an orbit and will provide a quick alert system for strong transients.
If we set the SRM bands limits at the energy where the probability that an event is assigned to each neighboring band is equal we find the following boundaries: 0.24, 0.38, 0.78, 1.54, 3.0, 6.0, 12, 24, 46, 91 MeV.

\section{Conclusions}

The mini-calorimeter of the AGILE mission is a versatile and powerful gamma-ray detector for the energy range 0.3 - 100~MeV based on segmented scintillator detectors with solid state readout. The inclusion of two parallel, simultaneously active acquisition branches allowed the full exploitation of the detectors capabilities, despite the strict constraint frame in which it has been developed. MCAL had to undergo several calibration and qualification steps before and after integration in the AGILE payload. The performance obtained on ground satisfy the scientific requirements.
AGILE was launched on April 23 2007 and is currently fully operative. After more than one year of in-orbit operations MCAL is fully functional and scientific exploitation of the MCAL data is ongoing.

\section{Acknowledgments}
AGILE is a mission of the Italian Space Agency, with co-participation of INAF (Istituto Nazionale di Astrofisica) and INFN (Istituto Nazionale di Fisica Nucleare). The authors wish to thank all the AGILE team for their help and fruitful discussions. The authors wish to thank also the industrial partners, namely the AGILE people at Thales-Alenia Space Italia and Carlo Gavazzi Space, for their fundamental contribution to the construction, integration and testing of MCAL. Many thanks also to Dr.~Elio Rossi, Alessandro Mauri and Alessandro Traci for their skillful electronics and mechanical prototyping and testing.




\appendix
\section{Energy and position reconstruction in an AGILE MCAL bar}
\label{reconstruction}
In this section the analytical algorithm used for energy and position reconstruction in an AGILE MCAL bar is described. Throughout this section a perfect exponential behavior in the light output of a bar is assumed, as described in subsection \ref{detection_plane}.

Let's consider a reference frame centered at the center of a bar having total length $L$, so that PD-A and PD-B are located at position $-\frac{L}{2}$ and $\frac{L}{2}$ respectively.

For a particle hitting the bar at a position $x$ and releasing energy $E$, the signal output $U_A$ and $U_B$ measured at the two PDs and expressed in ADC channels are:


	\begin{equation}
	\label{eq:UA}
U_A(x,E) = O_A + u_{0,A} E e^{-{\alpha}_A (x + \frac{L}{2})}
	\end{equation}


	\begin{equation}
	\label{eq:UB}
U_B(x,E) = O_B + u_{0,B} E e^{-{\alpha}_B (\frac{L}{2} - x)}
	\end{equation}

where:\\
$O_A$ and $O_B$ are the offsets of PD-A and PD-B electronic chains respectively, expressed in ADC channels;\\
$u_{0,A}$ and $u_{0,B}$ are the gain of PD-A and PD-B electronic chains respectively, expressed in ADC channels/MeV;\\
${\alpha}_A$ and ${\alpha}_B$ are side A and B light attenuation coefficients respectively, expressed in $cm^{-1}$.\\

If we define $\bar{\alpha} = \frac{{\alpha}_A + {\alpha}_B}{2}$ and $\Delta{\alpha} = \frac{{\alpha}_A - {\alpha}_B}{2}$ we obtain:


	\begin{equation}
	\label{eq:pos}
x = \frac{ln\frac{u_{0,A}}{u_{0,B}} - \Delta\alpha L}{2\bar{\alpha}} + \frac{1}{2\bar{\alpha}} ln\frac{U_B - O_B}{U_A - O_A}
	\end{equation}


	\begin{equation}
	\label{eq:energy}
E = \frac{e^{\bar{\alpha} \frac{L}{2}}}{\sqrt{u_{0,A} u_{0,B}}} e^{\Delta\alpha x} \sqrt{(U_A - O_A)(U_B - O_B)}
	\end{equation}

Neglecting the asymmetry in the light attenuation coefficients leads to the following simplified equations:


	\begin{equation}
	\label{eq:pos2}
x = \frac{ln\frac{u_{0,A}}{u_{0,B}}}{2\bar{\alpha}} + \frac{1}{2\bar{\alpha}} ln\frac{U_B - O_B}{U_A - O_A}
	\end{equation}


	\begin{equation}
	\label{eq:energy2}
E = \frac{e^{\bar{\alpha} \frac{L}{2}}}{\sqrt{u_{0,A} u_{0,B}}} \sqrt{(U_A - O_A)(U_B - O_B)}
	\end{equation}



\end{document}